\documentclass[twocolumn,showpacs,prl,amsfonts,amsmath,amssymb,floatfix,groupedaddress]{revtex4}
\usepackage{color}
\usepackage{hhline}
\usepackage{mathrsfs}
\usepackage{graphicx}
\usepackage{dcolumn}
\usepackage{bm}
\usepackage{multirow}
\usepackage{booktabs}
\usepackage{afterpage}
\usepackage{amsmath}
\usepackage{ulem}

\begin{document}

\title{Neural-network Kohn-Sham exchange-correlation potential and its out-of-training transferability}

\author{Ryo Nagai$^{1}$}
\author{Ryosuke Akashi$^{1}$}
\thanks{akashi@cms.phys.s.u-tokyo.ac.jp}
\author{Shu Sasaki$^{1}$}
\author{Shinji Tsuneyuki$^{1,2}$}
\affiliation{$^1$Department of Physics, The University of Tokyo, Hongo, Bunkyo-ku, Tokyo 113-0033, Japan}
\affiliation{$^2$Institute of Solid State Physics, The University of Tokyo, Kashiwa, Chiba 277-8581, Japan}

\date{\today}
\begin{abstract}
We incorporate in the Kohn-Sham self consistent equation a trained neural-network projection from the charge density distribution to the Hartree-exchange-correlation potential $n \rightarrow V_{\rm Hxc}$ for possible numerical approach to the exact Kohn-Sham scheme. The potential trained through a newly developed scheme enables us to evaluate the total energy without explicitly treating the formula of the exchange-correlation energy. With a case study of a simple model we show that the well-trained neural-network $V_{\rm Hxc}$ achieves accuracy for the charge density and total energy out of the model parameter range used for the training, indicating that the property of the elusive ideal functional form of $V_{\rm Hxc}$ can approximately be encapsulated by the machine-learning construction. We also exemplify a factor that crucially limits the transferability--the boundary in the model parameter space where the number of the one-particle bound states changes--and see that this is cured by setting the training parameter range across that boundary. The training scheme and insights from the model study apply to more general systems, opening a novel path to numerically efficient Kohn-Sham potential.
\end{abstract}

\maketitle

-{\it Introduction.}
The Kohn-Sham (KS) equation based on density functional theory~\cite{PhysRev.136.B864,PhysRev.140.A1133},
has long been the method of first choice for theoretical study on electronic and structural properties in atomic, molecular and bulk systems for its practical balance of numerical accuracy and cost. The general use of this scheme is formally justified by two facts: For an interacting electron system under arbitrary ionic potential $V_{\rm ion}$, (i) the ground state properties such as the total energy $E_{\rm tot}$ and response functions are unambiguously determined from the ground-state electron charge density distribution $n$ and (ii) the corresponding non-interacting Hamiltonian characterized by one-body potential $V_{\rm s}=V_{\rm ion}+V_{\rm Hxc}$ can be constructed so that the same $n$ is reproduced. The Hartree-exchange-correlation potential $V_{\rm Hxc}\equiv V_{\rm Hxc}[n]$, being the functional of $n$, is defined by the functional derivative of the Hartree-exchange-correlation energy: $V_{\rm Hxc}([n]; {\bf r})=\frac{\delta E_{\rm Hxc}}{\delta n({\bf r})}$. The exact form of $E_{\rm Hxc}$ is assured to be present and universal~\cite{PhysRev.140.A1133, Levy01121979}, but has been elusive.

Development of useful approximate forms for $E_{\rm Hxc}$ is mostly based on the common strategy, often referred to as ``Jacob's ladder"~\cite{Perdew-Ladder}, of systematic extension to include the dependence on the spatial derivatives of the density~\cite{PhysRev.136.B864} and KS orbitals~\cite{Perdew-Zunger1981, Becke-hybrid1993} (see Refs.~\onlinecite{Scuseria2005669,RevModPhys.80.3, Peverati20120476} for a review). Various handy and practically accurate functionals are in general use nowadays~\cite{Perdew-Zunger1981,PhysRevLett.77.3865,Becke-3param1993,Kim-Jordan-B3LYP, Stephens-Devlin-B3LYP,PBE0,PhysRevLett.91.146401,HSE03}. However, further extensions obviously suffers from severe obstacles. Increasing number of terms will eventually yield diverging options on how to determine the function forms and parameters entering $E_{\rm Hxc}$. Moreover, determination of the parameters referring to energy and related quantities, which is adopted in the construction of many functionals, sometimes yield inaccurate charge density~\cite{Medvedev49,Kepp496}. This could originate from the confusion of two types of error~\cite{PhysRevLett.111.073003}; error coming from the approximate form of $E_{\rm Hxc}$ itself, and that from the incorrect charge density yielded by the KS equation using the approximate $V_{\rm Hxc}$.

Overall, the systematic {\it formal} extension of $E_{\rm Hxc}$ has been a formidable task; in this paper, we explore an alternative {\it numerical} implementation of $V_{\rm Hxc}$, $E_{\rm Hxc}$ and self-consistent solution of the Kohn-Sham equation that can be systematically improved to the numerically exact limit. Apart from the unknown ideal form of $V_{\rm Hxc}[n]$, it is ultimately a projection $n \rightarrow V_{\rm Hxc}$; this fact allows us to implement it by a machine-learning technique. Wagner and coworkers have recently demonstrated a numerically exact KS scheme by iterative inverse derivations of the external potentials for interacting and non-interacting systems that reproduce the input $n$~\cite{PhysRevLett.111.093003, Burke-exact-xc2014}. Our proposal corresponds in spirit to the replacement of this inversion procedure with the machine-learning projection.

\begin{figure}[b!]
 \begin{center}
  \includegraphics[scale=0.7]{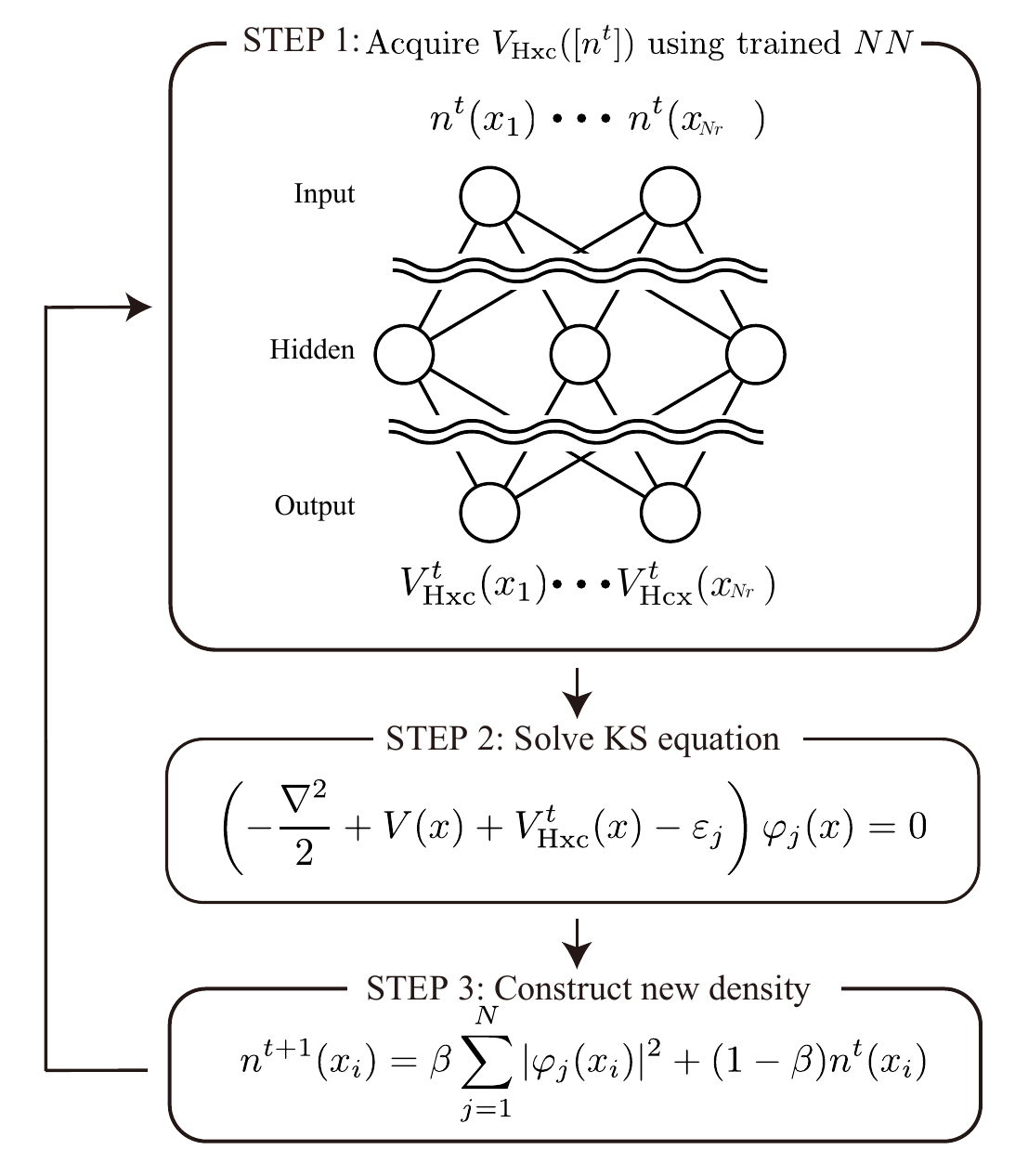}
  \caption{
  	Self-consistent KS cycle using the NN-$V_{\rm Hxc}$.
  	STEP 1: Substitute the charge distribution $n^{(t)} (x)$ to the trained NN to get $V_{\rm Hxc}$.
  	STEP 2: Solve KS equation using the $V_{\rm Hxc}$. 
  	STEP 3: Calculate new density $n^{(t + 1)} (x)$ and go back to STEP 1.
  	Repeat this cycle with $t\rightarrow t+1$until convergence.
  }
  \label{fig:graph-NN}
 \end{center}
\end{figure}

We construct the numerical projection $n \rightarrow V_{\rm Hxc}$ with the feed-forward neural network (NN) (Fig.~\ref{fig:graph-NN}), dubbed later as NN-$V_{\rm Hxc}$,
\begin{eqnarray}
{\bm V}_{\rm Hxc} = \cdots f[W^{(2)}f[W^{(1)}{\bm n} + {\bm b}^{(1)}]+{\bm b}^{(2)}] \cdots
.
\label{eq:VHxc-NN}
\end{eqnarray}
Here, ${\bm V}_{\rm Hxc}$ and ${\bm n}$ are any vectorized representations of $V_{\rm Hxc}$ and $n$ and $f$ is a non-linear activation function operating on each vector components. $W^{(l)} (l=1, 2,\dots)$ and ${\bm b}^{(l)}$ are the weight matrix and bias vector whose components are optimized to minimize the training error. Although there is no guarantee that the true functional form of $V_{\rm Hxc}$ is of Eq.~(\ref{eq:VHxc-NN}), its flexibility represented by the universal approximation theorem~\cite{HORNIK1991251} enables us to make it numerically indistinguishable from the true one with appropriate amount and quality of the training data. 

The NN-$V_{\rm Hxc}$ approach has several practical advantages. First, the strategy to the numerical accuracy is simple (increase the number of hidden layers, nodes, their connections, and training data sets). Second, in contrast to the conventional construction referring to the energy-related quantities, the density-driven part of the error~\cite{PhysRevLett.111.073003} is directly minimized. Third, it allows us self-consistent solution of the KS equation with computational costs of $O(N_{\rm r}^2)+O(N_{\rm r}^3)$ with $N_{\rm r}$ representing the system size; the former and latter are for calculating the potential and solving the KS equation, respectively, where the number of nodes per NN layer is presumably $O(N_{\rm r})$. This is smaller than the algorithm with the exact exchange operator [$O(N_{\rm r}^4)$].  

Note that the machine-learning technique can give us the $O(N_{r}^2)$ algorithms for the variational optimization of $n$ without solving the KS equation; e.g., by training the projections from $n$ to the universal energy functional~\cite{Burke-orbitalfree-PRL2012, PhysRevB.94.245129} and from $V_{\rm ion}$ to $n$~\cite{Burke-bypassing}. We find, on the other hand, that the explicit treatment of the kinetic energy operator in the KS equation, which obviously serve as a regulator of the spurious spatial oscillations in $n$, yield better accuracy out of the range of training data set at the expense of the larger computational cost. 

Our central interest is how the well-trained NN-$V_{\rm Hxc}$ is transferable: namely, if we construct a numerically accurate NN-$V_{\rm Hxc}$ in a certain space of matters, to what extent can it retain its accuracy out there? It seems plausible to expect the transferability, in an analogy to the fact that the LDA functional, which is exact only in the homogeneous limit, is practically accurate for inhomogeneous systems. We see below that the trained potentials indeed retain their accuracy comparable to the validation error well beyond the range of the models used for the training, implying that the machine-learning construction can grasp the character of the ideal functional.

Here we outline the procedures of the construction and test of the NN-$V_{\rm Hxc}$. (I) The learning data set $(n, V_{\rm Hxc})^{(i)} (i=1, 2, \dots, N_{\rm data})$ is generated with the two steps: for a certain potential $V^{(i)}_{\rm ion}$, (I-i) solve the interacting Hamiltonian with a numerically accurate solver to get the ground-state charge density, (I-ii) obtain $V_{\rm Hxc}^{(i)}$ as a solution of the inverse problem of reproducing $n^{(i)}$ via the KS equation. (II) Using the data set generated with $V_{\rm ion}$ within a certain parameter region, the NN-$V_{\rm Hxc}$ is trained. (III) Finally, using the trained $V_{\rm Hxc}$, we execute the KS self-consistent cycle (Fig.~\ref{fig:graph-NN}) for a given $V_{\rm ion}$ within and out of the learning range and compare the calculated $n$ and $E_{\rm tot}$ with those obtained by the accurate solver.

-{\it Energy level calibration for inverse problem.}
When one construct $V_{\rm Hxc}$ with the machine learning, there are two fundamental caveats. First, in generating the training $V_{\rm Hxc}$ through the solution of the inverse problem, it has an arbitrary constant term as it does not affect the wave function. Unless properly regularized, this obviously induce violent dependence of $V_{\rm Hxc}$ on $n$, reducing the training efficiency. Second, the total energy $E_{\rm tot}$ cannot be reproduced with only $V_{\rm Hxc}$; $E_{\rm Hxc}$ is needed as seen in the formula~\cite{PhysRev.140.A1133}
$E_{\rm tot}=\sum_{j}\varepsilon_{j}+E_{\rm Hxc}-2E_{\rm H}-\int d{\bf r} V_{\rm Hxc}({\bf r}) n({\bf r}).$
Here, $E_{\rm H}$ is the Hartree part of $E_{\rm Hxc}$. Numerical determination of $E_{\rm Hxc}$ so that it reproduces $V_{\rm Hxc} \equiv \frac{\delta E_{\rm Hxc}}{\delta n}$ is quite nontrivial task. 

The above two problems are simultaneously solved by utilizing the formulation by Levy and Zahariev~\cite{PhysRevLett.113.113002} setting the constant term of $V_{\rm Hxc}$ so that the total energy is reproduced by the sum of the energy eigenvalues of the KS equation: $E_{\rm tot}= \sum_{j} \varepsilon_{j}$. This means that we rewrite the above formula with a scalar functional $c\equiv c[n]$
\begin{eqnarray}
\hspace{-10pt}
E_{\rm tot}
\!=\!\!
\sum_{j} (\varepsilon_{j}\!+\!c)
\!+\!\!E_{\rm Hxc}
\!-\!2E_{\rm H}
\!-\!\! \! \int \!\!d{\bf r} (V_{\rm Hxc}({\bf r})\!\!+\!c) n({\bf r})
\label{eq:tot-ene2}
\end{eqnarray}
and adjust $c$ so that the second and later terms in Eq.~(\ref{eq:tot-ene2}) cancel. We are thus able to circumvent the problem of determining the functional form of $E_{\rm Hxc}$, as well as regularize the learning data in a well-defined and numerically stable way.

\begin{figure*}[t!]
 \begin{center}
  \includegraphics[scale=0.21]{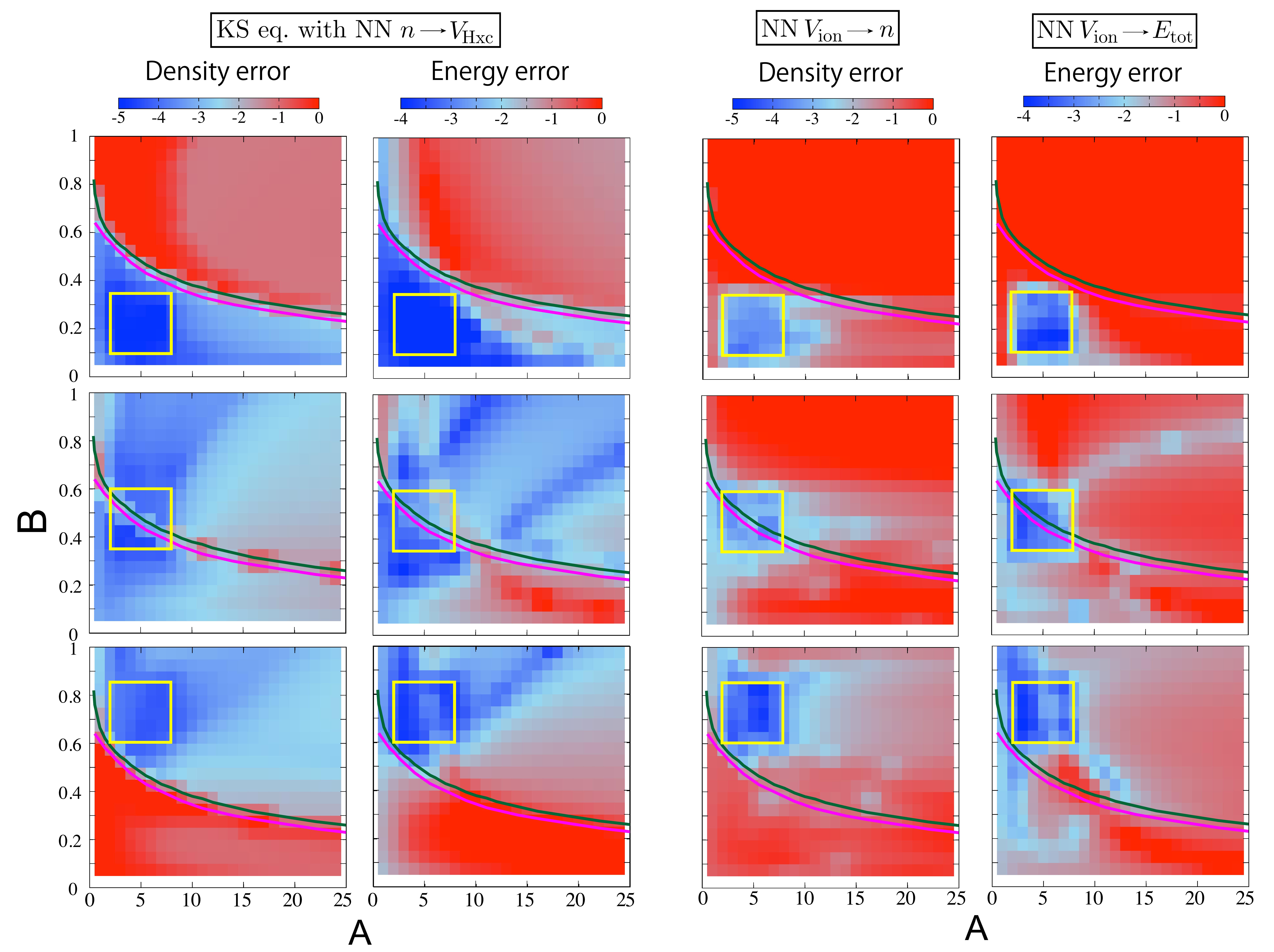}
  \caption{Transferability of NN-$V_{\rm Hxc}$ (first and second columns) and NN projections $V_{\rm ion} \rightarrow n$ (third) and $V_{\rm ion} \rightarrow E_{\rm tot}$ (fourth) trained within the parameter ranges indicated by the bold frames. The errors compared with the exact diagonalization results are plotted as color maps. Errors of density are ${\log_{10}}\sqrt{\Sigma_i(n(x_i)-n^{\rm exact}(x_i))^2/N_{\rm r}}$ and the errors of energy are $\log_{10}|(E_{\rm tot}-E_{\rm tot}^{\rm exact})/E_{\rm tot}^{\rm exact}|$. The lines show the trap boundaries with (green) and without (pink) the Coulomb interaction.}
  \label{fig:V-transfer}
 \end{center}
\end{figure*}

-{\it Model.}
We study a simple model system~\cite{note-interaction}: Two interacting spinless Fermions with an external Gaussian potential in one dimension ($x\in[-1, 1]$) with periodic boundary condition
\begin{eqnarray}
&&\left[
\sum_{k=1,2}
\bigg(
-\frac{\nabla_{k}^2}{2}
+
V(x_{k})
\bigg)
+\frac{1}{|x_{1}-x_{2}|}
\right]
\Psi
=E_{\rm tot}\Psi
\label{eq:Hamiltonian}
\\
&& V(x)=-A\exp\left(-x^2 /B^2\right).
\label{eq:potential}
\end{eqnarray}
$\Psi\equiv\Psi(x_1, x_2)$ is the two-particle wave function and the charge density is given by $n(x)=\int dx' |\Psi(x,x')|^2$. The corresponding Kohn-Sham equation reads
\begin{eqnarray}
\bigg(-\frac{\nabla^2}{2}+V(x)+V_{\rm Hxc}([n]; x)\bigg)\varphi_{j}(x)
=\varepsilon_{j}\varphi_{j}(x)
\end{eqnarray}
with $n(x)=\sum_{j} |\varphi_{j}(x)|^2$. This system is remarkable in that the number of one-particle bound states can be tuned by changing $A$ and/or $B$, which is a simplest modeling of the varying valence number of an ion. In the absence of the Coulomb interaction, when $A$ and $B$ are small, the energy eigenvalue of the one-body state is negative (bound) whereas that of the first excited state is positive (unbound). On the other hand, if $A$ or $B$ is large enough, the both states are bound. There is hence a boundary line in the $A-B$ space through which the distribution of the electron density dramatically changes, which we later refer to as ``trapping boundary". The boundary also applies to the interacting case, though its position slightly changes due to the Coulomb repulsion~\cite{suppl}.

Hereafter we treat all the functions of $x$ by the real-space discretization onto common $N_{r}$ uniform mesh points with $N_{r}=100$: NN-$V_{\rm Hxc}$ is then implemented as a projection from vector $\{n(x_{1}), n(x_{2}), \dots, n(x_{N_{r}})\}$ to vector $\{V_{\rm Hxc}(x_{1}), V_{\rm Hxc}(x_{2}), \dots, V_{\rm Hxc}(x_{N_{r}})\}$.

\begin{figure}[h!]
 \begin{center}
 \includegraphics[scale=0.20]{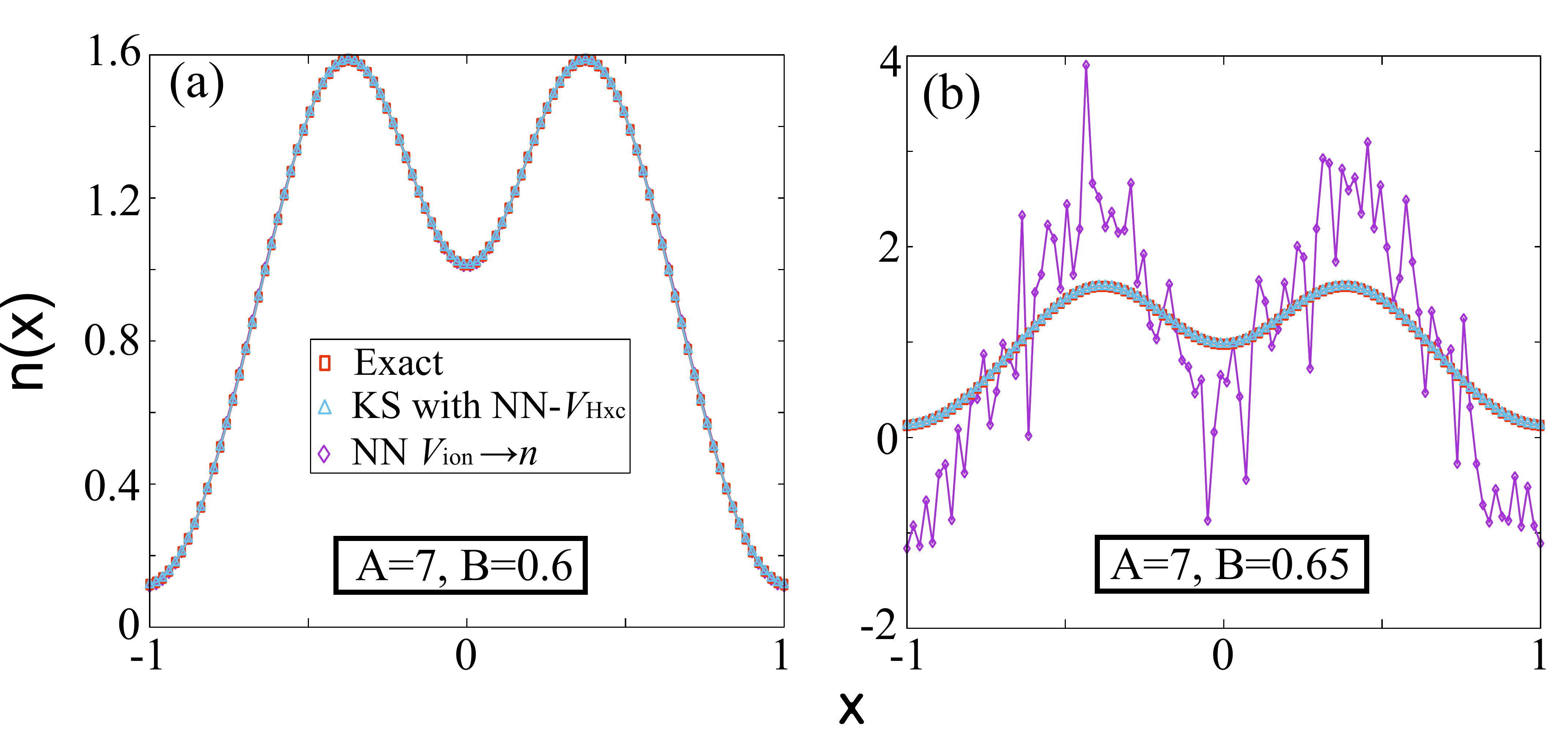}
  \caption{Typical charge density distributions obtained with the different methods. The training for the NNs has been done within Area II; $(A, B) \in [2.0, 8.0]\times [0.35, 0.60]$. (a) Results obtained for the setting $A=7.0$ and $B=0.60$ (within Area II) and (b) $A=7.0$ and $B=0.65$ (out of Area II), respectively.}
  \label{fig:charge-oscillation}
 \end{center}
\end{figure}

-{\it Generation of training data.}
To examine the effect of the boundary on the transferability of the NN-$V_{\rm Hxc}$, we prepared three distinct training data sets yielded by the potentials respectively generated in the parameter $(A, B)$ regions I$=[2.0, 8.0]\times[0.10, 0.35]$, II$=[2.0, 8.0]\times[0.35, 0.60]$, and III$=[2.0, 8.0]\times[0.60, 0.85]$. These regions are designed so that they are within, crossing, and totally beyond the boundary line, respectively (Fig.~\ref{fig:V-transfer}). Exact diagonalization was employed as the accurate solver. The specific procedure to generate the training data sets with the above-mentioned energy level calibration is appended in Supplemental Materials~\cite{suppl}. For the inverse Kohn-Sham step, we adopted the method~\cite{HF} based on the Haydock-Foulkes variational principle~\cite{variational}.

-{\it Training and Kohn-Sham cycle.}
The generated data sets $(n, V_{\rm Hxc})^{(i)} (i=1,2,\dots, N_{\rm data})$ were used for optimizing the parameters $W$ and ${\bm b}$ of the NN. We used the fully-connected NN with two hidden layers having 300 nodes. The rectified linear unit~\cite{ReLU} was used as the activation function $f$. The root mean squared error between $V_{\rm Hxc}^{(i)}$ and those calculated from $n^{(i)}$ by NN was minimized with the adaptive moment estimation method (AdaM~\cite{Adam}). Details of the optimization and typical behavior of the error are appended in Supplemental Materials~\cite{suppl}.

Using the trained NN-$V_{\rm Hxc}$, the Kohn-Sham self-consistent cycle (Fig.~\ref{fig:graph-NN}) was executed with the linear mixing parameter $\beta=0.4$. The cycle was repeated until the convergence condition $\sqrt{\sum_{i}|n^{t+1} (x_i) - n^{t} (x_i)|^2/N_{r}} < \sigma$ was achieved with $\sigma = 5\times10^{-8}$, which was set equal to the convergence threshold for the inverse KS procedure~\cite{suppl}. \begin{figure}[t]
 \begin{center}
  \includegraphics[scale=0.20]{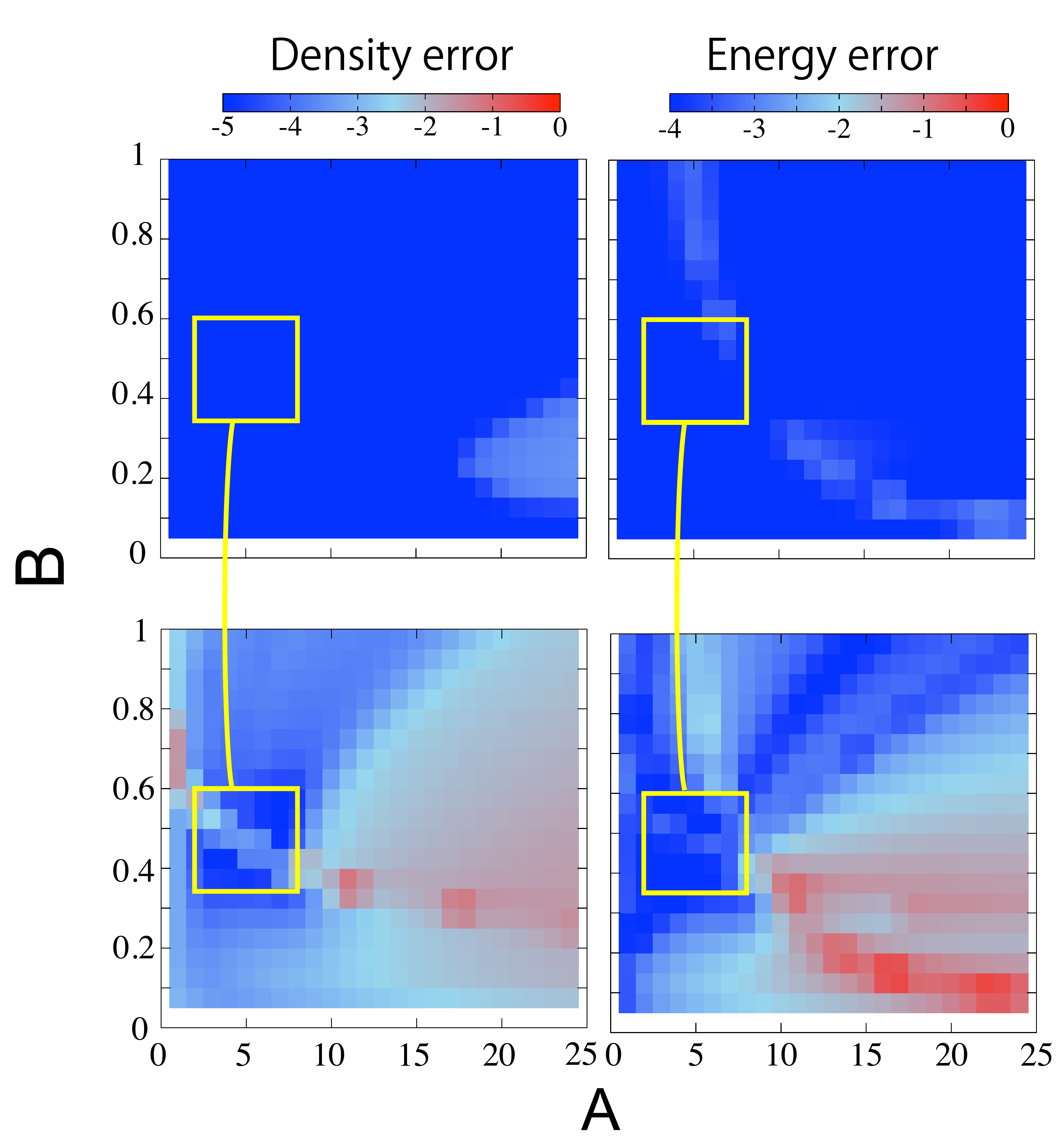}
  \caption{Transferability of NN-$V_{\rm Hxc}$ trained with data sets including the cases $\int dx n(x)=1$ and $2$, where box frames denote the training parameter range. The upper (lower) panels display the errors for the calculations with $\int dx n(x)=1$ (2), compared with the exact diagonalization results as in Fig.~\ref{fig:V-transfer}.}
  \label{fig:V-transfer-1-2}
 \end{center}
\end{figure}

-{\it Results.}
The left two columns in Fig.\ref{fig:V-transfer} show the out-of-training error in $n$ and $E_{\rm tot}$ derived with the KS scheme using the NN-$V_{\rm Hxc}$ with the training parameter ranges I--III drawn as the box frames. The errors are defined by the difference from the exact diagonalization results $n^{\rm exact}$ and $E^{\rm exact}_{\rm tot}$. Remarkably, the NN-$V_{\rm Hxc}$ method reproduced the two-body density and energy with good accuracy up to far out of the training ranges, indicating its transferability. Another notable thing is that the trapping boundary crucially affects the accuracy of the trained NN-$V_{\rm Hxc}$. When the training range is taken wholly within one side from the boundary (upper or lower row), the trained NN-$V_{\rm Hxc}$ becomes inaccurate across the boundary. This reflects that the character of the exact charge density distribution varies significantly. When the training range is Area II, which is crossing the boundary (middle row), the transferability is retained for the whole parameter range displayed, though there still remains a trend of low accuracy near the boundary. 

To see the effect of explicit treatment of the kinetic energy operator, we also trained two types of ``bypassing~\cite{Burke-bypassing}" projections by the NN: $V_{\rm ion} \rightarrow n$ and $V_{\rm ion} \rightarrow E_{\rm tot}$ (Fig.~\ref{fig:V-transfer}, right columns). Although we achieved the accuracy comparable to the present KS scheme with the NN-$V_{\rm Hxc}$ within the parameter ranges for training, the out-of-training transferability is apparently worse. For the parameters out of the training range, we frequently observed spatially oscillating charge density distributions as reported in Ref.~\onlinecite{Burke-orbitalfree-PRL2012}; this behavior is suppressed with the present method, demonstrating the effect of the kinetic energy operator as a regulator of the artificial oscillation~(Fig.~\ref{fig:charge-oscillation}). 


We thus clarified the advantage of keeping the $O(N^{3}_{r})$ KS formalism. Another advantage is that, with the projection of form $n\rightarrow V_{\rm Hxc}$, it can formally treat the systems having different number of electrons with the unified formula of $V_{\rm Hxc}$. To demonstrate this, we trained the NN-$V_{\rm Hxc}$ using the training data with its total number of electrons 1 and 2. The parameter range for $V_{\rm ion}$ was set to II~\cite{note-learn-N12}. Note that this includes the most pathological case; $V_{\rm Hxc}$ must be zero for any $n$ when $\int$ $dxn(x)=1$. The resulting accuracy for the both cases is commonly high, as shown in Fig.\ref{fig:V-transfer-1-2}.

-{\it Summary and conclusions.}
We have studied the performance of the KS cycle incorporating the NN-$V_{\rm Hxc}$. A convenient scheme for its training has been developed, which enables us to get the total energy without resorting to explicit forms of $E_{\rm Hxc}$, as well as regulate the arbitrary constant in the KS inversion procedure. For the spinless two-electron model, the present method has been shown to reproduce both the density and total energy within the training area, as well as far out of it. The trapping boundary, where the character of the charge density varies from localized to itinerant, has been found to restrict the transferability. This fact gives us insight for general use of the machine-learning $V_{\rm Hxc}$, especially for multivalence systems such as Fe$^{2+}$ and Fe$^{3+}$; the range of training data in the compound space must cover the cases with various valence numbers.

Provided that the training proceeds efficiently, the scheme of implementing NN-$V_{\rm Hxc}$ itself does not contain any physical approximations except those in the training data. Also, we have seen that by retaining the KS-equation formalism the trained potential acquire remarkable transferability, which suggests that the amount of the training data required for general practice is somehow small. The Kohn-Sham scheme with a trained NN-$V_{\rm Hxc}$ could hence pave a way to the universal calculation method that brings various systems into the target, including the strongly correlated systems.

\begin{acknowledgments}
-{\it Acknowledgment.}
R.~A. thanks to Kieron Burke and Yoshihide Yoshimoto for enlightening comments.
\end{acknowledgments}

\bibliography{reference}

\begin{thebibliography}{33}
\bibitem[{not({\natexlab{a}})}]{note-interaction}
\bibinfo{note}{The long range part of the interaction was truncated within
  $|x_{1}-x_{2}| < 2$, whereas we approximated $1/|x_{1}-x_{2}|$ by a large
  value at the singularity $x_{1}=x_{2}$.}
\bibitem[{not({\natexlab{b}})}]{note-learn-N12}
\bibinfo{note}{Two sets of 1000 training data were generated for $N_{\rm
  tot}=$1 and 2.}
\end{thebibliography}
\clearpage

\section{Supplemental Materials}
\subsection{The inverse KS procedure with the energy-level calibration}
Here we explain the procedure of generating the training data through the Kohn-Sham inversion with the energy level calibration.
\begin{enumerate}
 \item Take the parameters $(A, B)$ in the external potential $V_{\rm ext}$ randomly.
 \item Solve the two-body Schr\"odinger equation for Hamiltonian Eq.~(2) to get the total energy $E_{\rm tot}$ and the electron density $n(x)$.
 \item Set $V_{\rm Hxc}^0(x_i)=0$ for all $i$.
 \item Solve the KS equation with the one-body potential $V_{\rm ext}+V_{\rm Hxc}^{t}$ with the exact diagonalization to get the KS energy $\{\varepsilon_{j}\}$ and the density $n^t(x)$
 	 \begin{eqnarray}
 		&&H=\Bigl(-\frac{\nabla^2}{2}+V_{\rm ext}(r)+V_{\rm Hxc}^{t}(r)\Bigr)\\
		&&n^t(x_i)=\sum_{j=1}^N |\varphi_{j}(x_i)|^2
 	\end{eqnarray}
 \item 	For each $x_{i}$, update the $V_{\rm Hxc}^{\rm t}(x_i)$ by
 	\begin{eqnarray}
		V_{\rm Hxc}^{\rm t+1}(x_i)=V_{\rm Hxc}^{\rm t}(x_i)+\alpha(n^t(x_i)-n(x_i))
	\label{eq:updatingV_Hxc}
	\end{eqnarray}
	The mixing parameter $\alpha$ was set to $450$.
\item Repeat the steps 4--5 until $\sqrt{\sum^{N_{r}}_{i}|n^{t+1} (x_i) - n^{t} (x_i)|^2/N} <s$ is achieved. The convergence threshold $s$ was set to $5\times10^{-8}$ in the present work.
\item Adjust the constant part of the converged $V_{\rm Hxc}(x_i)$
	\begin{eqnarray}
 		V_{\rm Hxc}\rightarrow V_{\rm Hxc}+\frac{1}{N}(E_{\rm tot}-\sum_{j=1}^N\varepsilon_{j})
 	\end{eqnarray}
	so as to make the sum of KS energy $\sum_{i=1}^N\varepsilon_i^{\rm new}$ solved under the adjusted $V_{\rm Hxc}$ equates to $E_{\rm tot}$ (adjust $c$ in Eq.~(2) so that the second and later terms cancel).
	\begin{eqnarray}
 		\sum_{j=1}^N\varepsilon_{j}^{\rm new} &=&
		\sum_{j=1}^N\{\varepsilon_{j}^{\rm old}+\frac{1}{N}(E_{\rm tot}-\sum_{j'=1}^N\varepsilon_{j'}^{\rm old})\}\nonumber\\
		&=&E_{\rm tot}
 	\end{eqnarray}
\end{enumerate}

\section{Charge density distributions for the model}
\begin{figure}[b!]
 \begin{center}
 \includegraphics[scale=0.25]{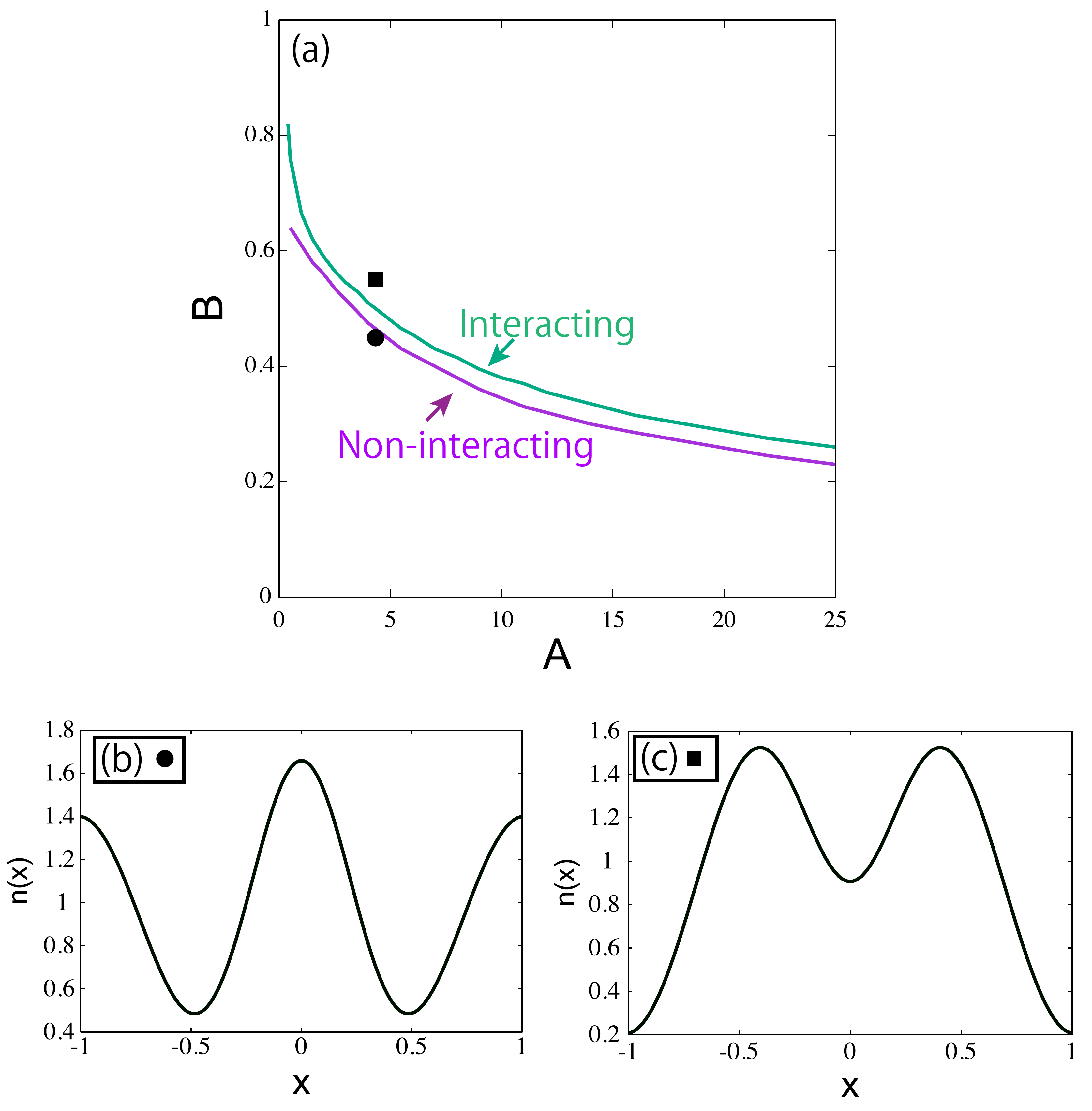}
  \caption{(a) Boundary line where the character of $n(x)$ for Hamiltonian Eq.~(3) changes. The line for the non-interacting case (without the interaction term) is rigorously defined as the line across which the number of the one-particle bound states changes from 1 to 2, whereas that for the interacting case was drawn by observing the rapidly changing behavior of $n(x)$ derived from the two-particle wave function. (b) Snapshot of $n(x)$ calculated at the point $(A, B)=(4.0, 0.45)$ indicated by circle in panel (a). (c) Snapshot of $n(x)$ calculated at the point $(A, B)=(4.0, 0.55)$ indicated by square in panel (a). }
  \label{fig:solution-trap}
 \end{center}
\end{figure}

The charge density realized by the Hamiltonian Eq.~(2) shows rapid change across the line in the A-B plane. As discussed in the main text, this reflects the change of the number of one-particle bound states in the non-interacting case. In Fig.~\ref{fig:solution-trap}, we here show a typical behavior of $n$ across the boundary for the interacting case. The boundary lines for the interacting case (Fig.~3) were drawn on the basis of this rapid change.

\begin{figure}[b!]
 \begin{center}
  \includegraphics[scale=0.26]{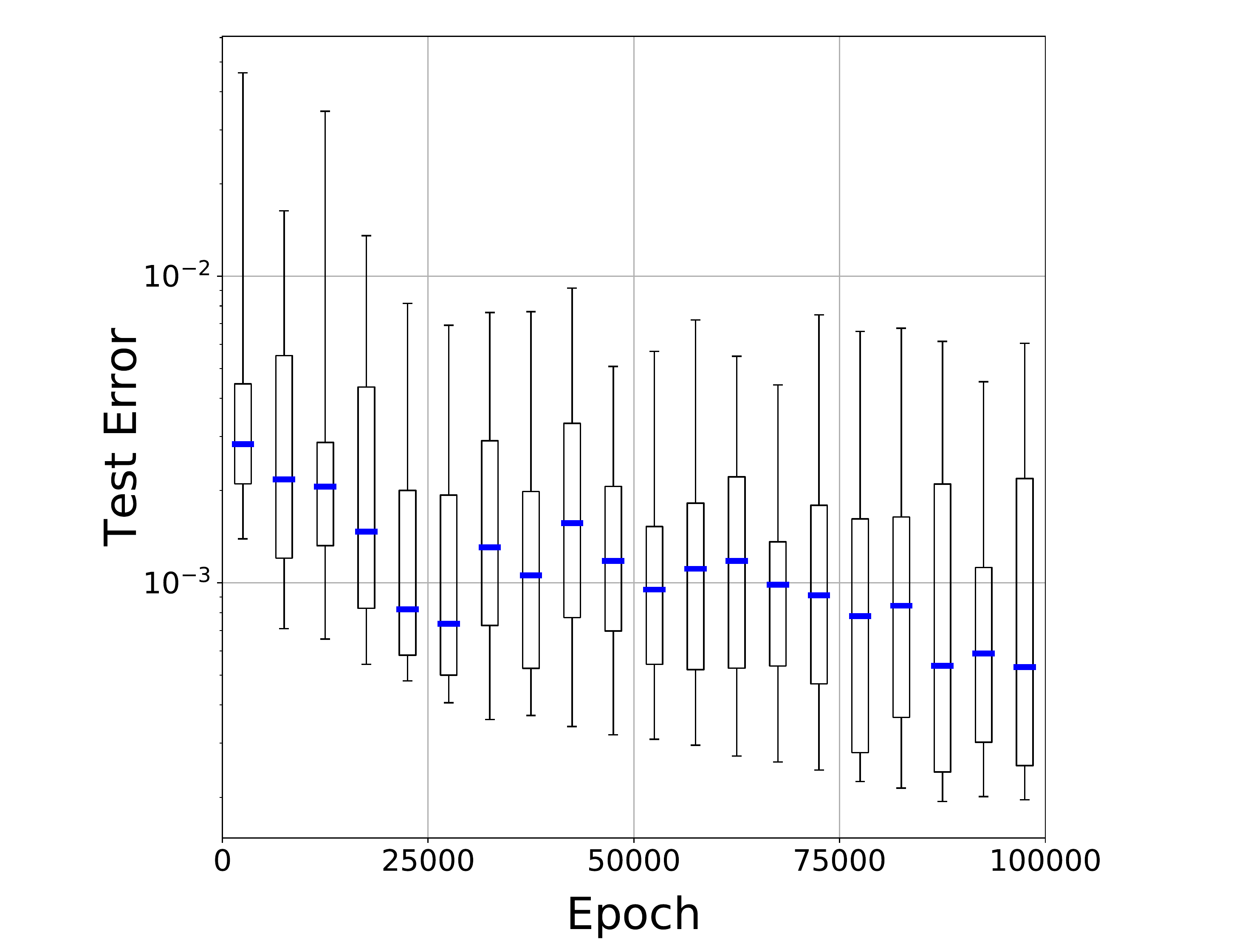}
  \caption{Typical learning curve of the training of NN-$V_{\rm Hxc}$ for the training data set I (see the main text). Each box represents the distribution of the test error for every 5,000 epochs. Box indicates the interquartile range, whiskers maximum and minimum values, and bold line in the box median of each 5,000 epochs.
  }
  \label{fig:learningline}
 \end{center}
\end{figure}

\section{Optimization of the neural network}
The initial step size of the AdaM was set to 0.001. Randomly selected 95\% of the generated data were referred to for training and the remaining 5\% were used to evaluate the test error. Figure \ref{fig:learningline} shows the typical behavior of the test error during the optimization. Since the test error does not reduce monotonically with the AdaM, the error distribution for every 5000 epoch is represented by the box plot. In this study we stopped the optimizations after 100,000 steps, though we have found that the stable convergence of the KS cycle and the trend of transferability shown in the main text are robust against at which step the optimization is stopped.

\end{document}